# Long-range magnetic perturbations at $Bi_2Te_3$/$Cr_2Te_3$ interfaces induced by chemical diffusion and proximity effects

Markel Pardo-Almanza[1], Yuita Fujisawa[1,2], Takatsugu Onishi[1], Chia Hsiu Hsu[1], Alec P. LaGrow[1], and Yoshinori Okada[1]

[1]Okinawa Institute of Science and Technology, Okinawa 904-0495, Japan

[2]Research Institute for Synchrotron Radiation Science, Hiroshima University, Higashi-Hiroshima, 739-0046, Japan

The heterointerface between topological insulators and magnetic materials provides a crucial platform for investigating exotic electronic and magnetic phases, with implications for both fundamental studies and potential applications. A key challenge is determining the spatial extent of magnetic perturbation across the interface. In this study, we grew $Bi_2Te_3$ films with mixed thicknesses of 5 and 6 quintuple layers (QLs) on a thick ferromagnetic $Cr_2Te_3$ film. Chemical analysis indicates gradual Cr diffusion across the interface into $Bi_2Te_3$, reaching up to the $4^{th}$ QL, while the $5^{th}$ and $6^{th}$ QL remain free of Cr. Despite the absence of direct Cr incorporation, quasiparticle interference imaging using spectroscopic scanning tunneling microscopy reveals strong backscattering on the $5^{th}$ QL surface. This suggests an intrinsic magnetic proximity effect, which breaks time-reversal symmetry without magnetic doping, extending into the $5^{th}$ QLs. Although variations may exist across different systems, the large magnetic perturbation length scale observed here is a valuable guiding principle for engineering exotic electronic and magnetic states at heterointerfaces between topological insulators and magnetic materials.

# I. INTRODUCTION

Heterostructures that combine materials with strong spin-orbit coupling (SOC) and magnetic order have emerged as promising platforms. The High SOC effect enables extensive control over magnetic and electronic states at the heterointerface, offering exciting opportunities for both fundamental studies in quantum materials and practical device applications [1,2,3]. However, accurately determining the length scales of interactions and symmetry breaking across the interface remains an experimental challenge. Interfacial phenomena are further enriched by chemical diffusion, which can drive novel emergent effects rather than merely posing an obstacle [4]. Achieving precise atomic-level control and a deep understanding of this complexity are essential for realizing desirable atomic- and nanoscale phenomena.

Among the relevant heterostructures, interfaces between $Z_2$ topological insulators (TIs) and magnetic materials are especially significant [2]. The $Z_2$ TI usually hosts a high SOC effect, resulting in topologically protected surface states (TSS) with spin-momentum locking [5,6,7,8]. Breaking time-reversal symmetry (TRS) in the TSS opens a gap at the Dirac point (DP), leading to the realization of novel states of matter, such as the quantum anomalous Hall insulator [9] and axion insulator states [10]. In contrast to the TRS breaking introduced through magnetic chemical doping [11], the magnetic proximity effect (MPE) in heterostructures offers an attractive approach that mitigates the introduction of chemical disorder [12,13,14,15]. Considerable efforts have been made to quantify the spatial extent of the MPE [12,13,14,15], though a clear consensus on its precise length scale remains elusive. Recent work highlights the critical role of chemical intermixing across interfaces [16], underscoring the importance of combining cross-sectional chemical characterization with surface-sensitive electronic state investigations to understand the effective magnetic perturbation length.

This study focuses on the heterostructure combining $Bi_2Te_3$ and the magnetic chalcogenide $Cr_2Te_3$. $Bi_2Te_3$ is a prototypical TI featuring a single Dirac cone at Γ point [**Fig. 1(a)**] [17]. Concurrently, self-intercalated $Cr_{1+\delta}Te_2$ films provide a tunable magnetic platform where the Curie temperature ($T_C$) and magnetic anisotropy can be systematically controlled by varying δ [18,19,20]. Specifically, $Cr_2Te_3$ (δ ~ 0.33) exhibits ferromagnetism ($T_C$ ~ 150 K) with strong perpendicular magnetic anisotropy. The structural and compositional compatibility between $Bi_2Te_3$ and $Cr_2Te_3$ make this an ideal platform for studying the magnetic perturbation length scale. Here, utilizing the characteristic heterostructure that exhibits the coexistence of regions with 5 and 6 QLs of $Bi_2Te_3$ grown on $Cr_2Te_3$, significant differences in the surface electronic states of these regions were observed using *in-situ* spectroscopic imaging scanning tunneling microscopy. Our findings suggest that the long-range magnetic perturbation arises from both magnetic chemical diffusion and the MPE.

## II. METHODS

The $Bi_2Te_3$/$Cr_2Te_3$ heterostructures [see **Fig. 1(a)**] were grown using molecular beam epitaxy (MBE), with simultaneous monitoring of the reflection high-energy electron diffraction (RHEED) pattern. First, 80 nm thick $Cr_2Te_3$ films were deposited on $Al_2O_3$ (001) substrates following the procedure described in [18,19]. After $Cr_2Te_3$, $Bi_2Te_3$ films were grown in the same MBE chamber. Further details on sample preparation are shown in **Supplementary Note 1** in the Supplemental Material (SM) [21]. The scanning tunneling microscopy/spectroscopy (STM/S) measurements were performed *in situ* at 4 K using a tungsten tip. For the *ex-situ* cross-sectional scanning transmission electron microscopy (STEM) measurements, a sample was capped *in-situ* with ~50 nm of fullerene to minimize surface degradation during the focused ion beam (FIB) lamella preparation. After the FIB preparation, the JEOL JEM-ARM200F operating at 200 kV was used to characterize crystal structure and chemical diffusion across the interface.

## III. EXPERIMENTAL RESULTS and DISCUSSIONS

**Length scale of Cr diffusion across the interface**

Using STEM energy-dispersive X-ray spectroscopy (EDS), we first evaluated the Cr diffusion length scale in the heterostructure [**Figs. 1(b)-(c)**]. Within the limits of experimental resolution, the analysis reveals that Cr diffusion is confined to approximately 4 nm from the interface (see **Supplementary Note 2** in [21] for the details of the cross-sectional chemical analysis). Given that 1 nm corresponds to roughly one quintuple layer (QL) thickness, Cr diffusion is restricted to $4^{th}$ QLs [see **Fig. 1(c)**]. As a result, the surface of the $5^{th}$ $Bi_2Te_3$ film remains unaffected by Cr diffusion. Therefore, it is crucial to investigate the signature of magnetic perturbation on the surface electronic structure of $5^{th}$ QL and thicker. This corresponding thickness is highlighted in **Fig. 1(c)**.

**STM characterization (surface point of view)**

**Figure 2(a)** presents an STM topographic image from the surface of $Bi_2Te_3$/$Cr_2Te_3$ heterostructure analyzed in this study. The hexagonal atomic pattern of the topmost Te layer in $Bi_2Te_3$ is clearly visible, as confirmed by the Bragg peaks in the Fourier-transformed (FT) image [circles in **Fig. 2(b)**]. The surface exhibits a pronounced topographic corrugation with a height variation of ~ 4 Å, as shown in the line profile in **Fig. 2(c)**. This ~ 4 Å corrugation can be explained by the mismatch in out-of-plane lattice parameters between $Cr_2Te_3$ (~ 6 Å) and $Bi_2Te_3$ (~ 10 Å) [see **Fig. 1(a)**]. When $Bi_2Te_3$ grows from an atomic step on the $Cr_2Te_3$ surface, the difference in QL thickness leads to height variations of ~ 4 Å, corresponding to the step height between adjacent quintuple layers [height differences between the $n^{th}$ and $(n+1)^{th}$ QLs are depicted in **Fig. 2(d)**]. Indeed, cross-sectional STEM imaging [**Fig. 2(e)**] directly confirms the coexistence of 5$^{th}$ QL and 6$^{th}$ QL regions in Bi$_2$Te$_3$, supporting the observed ~4 Å topographic corrugation.

Despite the pronounced topographic corrugation, the absence of atomic discontinuity on the topmost Te atomic sheet is notable [**Fig. 2(e)**]. This continuity is further corroborated by the high-pass-filtered topographic image [**Fig. 2(f)**], where a smooth connection is consistently observed between the 5$^{th}$ QL and 6$^{th}$ QL regions [pink rectangle area in **Fig. 2(a)**]. Such atomic continuity likely arises from the relatively facile lateral bonding characteristic of van der Waals (vdW)-coupled materials during film growth, forming a continuous Te atomic sheet rather than a discontinuous topmost layer [22]. For clarity, we designate areas with 5$^{th}$ QL and 6$^{th}$ QL as S$_5$ (lower regions) and S$_6$ (higher regions), respectively [see upper part in **Fig. 2(d)-(e)**]. It is worth noting that the STEM images in **Fig. 1(b)** and **Fig. 2(e)** originate from different locations of a film, while the same area as the STM image in **Fig. 2(a)** will be discussed hereafter.

**Heterogeneous bulk states ($E_{BVB}$ and $E_{BCB}$)**

**Fig. 3(a)** displays the averaged d$I$/d$V$ spectrum over the area shown in **Fig. 2(a)**, where two characteristic peaks can be assigned: the bulk valence band ($E_{BVB}$) and the bottom of the bulk conduction band ($E_{BCB}$). We further evaluate the spatial evolution of $E_{BVB}$ and $E_{BCB}$ from point spectra, as shown in **Figs. 3(b)** and **3(c)**, respectively (see **Supplementary Note 3** for the detailed procedure to obtain the characteristic energies in [21]). They show a similar contrast to the STM topographic image [**Fig. 2(a)**], suggesting that the origin of the variation exists in the different thicknesses of Bi$_2$Te$_3$. Consistent with this picture, the histograms of $E_{BVB}$ and $E_{BCB}$ show a bimodal trend for two energy scales [**Figs. 3(d)-(e)**]. To further clarify the underlying correlation, $E_{BVB}(r)$ versus $E_{BCB}(r)$ from all locations are plotted in **Fig. 3(f)**. A reasonable proportionality is recognized with the relation $E_{BVB}(r) - E_{BCB}(r) \sim 250$ meV, as indicated by the pink line in **Fig. 3(f)**. This means that the local bulk band gap size is locked at any location as ~ 250 meV, which is consistent with bulk band gap size ubiquitously observed in spectroscopic studies of Bi$_2$Te$_3$ [6]. The spatial variation of the bulk band shift ($\delta E_B$) reaches up to ~ 60 meV [see **Figs. 3(d)-(e)**], which is much larger than the conventional inhomogeneity from native defects/impurities or ripples seen in bulk [23,24,25]. Therefore, while the detailed mechanism is not fully cleared at this stage, the distinct diffused Cr density in the subsurface between the S$_5$ and S$_6$ regions is a reasonable hypothesis as an origin [26].

**Quasiparticle interference (QPI) pattern**

Through the analysis of the QPI pattern, the momentum, spin, energy, and spatial evolution of the TSS band can be investigated [25,27,28,29]. We first define relevant scattering channels based on spatially averaged QPI patterns at characteristic two energies (+ 50 and - 20 meV). The d$I$/d$V$ mappings and their Fourier-transformed (FT) images at two energies are displayed in **Figs. 4(a)-(d)**. The raw FT images [left panels in **Figs. 4(c)-(d)**] and their six-fold symmetrized versions [right panels in **Figs. 4(c)-(d)**] highlight the underlying

QPI scattering channels. In $Bi_2Te_3$, the bulk band potential induces energy-dependent hexagonal warping at energies far from the Dirac point (DP) [30]. In this energy region, the scattering vector along the Γ–M direction ($q_{ΓM}$) connects relatively parallel portions of the constant energy contour (CEC) [pink arrows in **Fig. 4(e)**]. The corresponding $q_{ΓM}$ is experimentally observed at + 50 meV [pink arrow in **Fig. 4(c)**], which is reasonable due to favorable nesting conditions rather than backscattering. However, if TRS is broken, a change of constant energy contour geometry with decreasing energy closer to the DP allows backscattering along the Γ–K direction [green arrows in **Fig. 4(f)**] [31,32]. Strikingly, the relevant backscattering $q_{ΓK}$ is experimentally observed at - 20 meV [green arrow in **Fig. 4(d)**]. To visualize the relative QPI intensity along Γ–K between the $S_5$ and $S_6$ regions, the terrace is divided into two areas (corresponding to $S_5$ and $S_6$) with equal surface fractions [**Fig. 2(a)** and **Fig. 3(b)-(c)**]. Analogous extraction of the spatial evolution of QPI patterns has been proven to be a powerful tool for investigating TIs [25,33].

**Energy evolution of QPI**

The QPI intensity profiles for the $S_5$ and $S_6$ regions are shown in **Figs. 5(a), (c)** and **Figs. 5(b), (d)**, respectively. When peak positions are resolved within the experimental resolution, dispersive QPI channels are quantitatively estimated and marked by the unfilled symbol for the right axis of **Figs. 5(a)-(d)**. For the $S_5$ region, a dispersive feature along the Γ–M direction is recognized [**Figs. 5(a)**] between + 100 and - 30 meV. For the $S_6$ region, such dispersive feature along Γ–M direction is similarly recognized between + 60 and - 90 meV [**Figs. 5(b)**]. On the other hand, QPI along Γ–K direction shows a striking contrast between the $S_5$ and $S_6$ regions across a broad energy range [**Figs. 5(c)-(d)**]. While dispersion is evident in the $S_5$ region [circles in **Fig. 5(c)**], such dispersion is absent in the $S_6$ region **[Fig. 5(d)]**.

A more rigorous comparison of QPI dispersion between the $S_5$ and $S_6$ regions requires accounting for energy shifts, given that potential difference is seen in the bulk band [**Fig. 3**]. The QPI dispersion from two regions is summarized in **Fig. 5(e)**, where the right and left axes represent energy for $S_5$ and $S_6$, respectively. As shown in **Fig. 5(e)**, aligning the dispersion along Γ-K between $S_6$ and $S_5$ yields an estimated energy shift of $δE_S$ ~ 45 meV. This energy scale is comparable to $δE_B$ ~ 60 meV, estimated as the bulk potential shift [**Figs. 3(d)–(f)**]. Considering the energy shift $δE_S$ ~ 45 meV, the QPI patterns at - 20 meV ($S_5$, red) and - 65 meV ($S_6$, blue) are fairly compared in **Fig. 5(f)**. Dividing the $S_5$ QPI image by $S_6$ maximizes the visibility of $q_{ΓK}$ in the $S_5$ region, with suppressing undesired background signal [**green circles in Fig. 5(g)**]. Also, a line cut along Γ–K further reveals a peak structure corresponding to the backscattering channel $q_{ΓK}$ [black arrow in **Fig. 5(h)**]. Since dispersion along Γ–K corresponds to backscattering indicated by $q_{ΓK}$ [see **Fig. 4(f)**], the signature of the TRS breaking is evident in the $S_5$ region, while the corresponding signal is missing in the $S_6$ region.

**Gap opening at the charge neutral point**

Although TRS breaking allows backscattering, it becomes difficult to detect near the charge neutral point because the joint density of states is lower for the circular CEC than for the hexagonal one [**Fig. 4(f)-(g)**]. On the other hand, the shape of the d$I$/d$V$ spectrum provides significant insight into the band structure near the charge-neutral point. In addition to minimum conductance [indicated by $E_{min}$ (~ -200 meV) in **Fig. 6(a)**], the d$I$/d$V$ spectrum from the $S_6$ shows only a shoulder at -150 meV ($E_w$). Since $E_w$ matches well with the characteristic energy scale above which the QPI signal along Γ-M becomes clear, the origin of $E_w$ is group velocity deformation of the surface state band driven by the hexagonal warping effect [see **Fig. 4(e)**]. The notable point is missing d$I$/d$V$ signature for the gap opening around $E_{min}$ in the $S_6$ region. In contrast, the d$I$/d$V$ spectrum from the $S_5$ region [**Fig. 6(b)**] shows qualitative distinct features compared to the $S_6$ region. The minimum conductance structure in the $S_5$ region is flanked by two humps around - 90 meV and -150 meV. These three adjacent characteristic energy scales, including one minimum ($E_{min}$) and two shoulder structures ($E_{gap+}$ and $E_{gap-}$), can be interpreted as gap formation around the charge neutral point of the surface state in $S_5$ [34]. This interpretation is consistently supported by rough extrapolation of QPI dispersion along Γ–K to $q_{ΓK}$ = 0 limit [**Fig. 6(c)**], based on a high-energy linear (dashed line) portion and a low-energy curved portion (solid line). Note that the horizontal axis of **Fig. 6(c)** corresponds to momentum along Γ–K direction, estimated by $k_{ΓK} = q_{ΓK}/2$.

By reducing the thickness of three-dimensional topological insulators, the properties of the surface states change significantly. Consequently, careful consideration is required to interpret the nature of the gap observed in the $S_5$ region. One possibility is that the gap arises from the mass term imparted to Dirac electrons [**Fig. 6(d)**], resulting in a gap opening at the Dirac point of a single Dirac cone in the $S_5$ region. Another possibility is the gap opening at the band crossing point on the Rashba-type band [**Fig. 6(e)**], which is initially driven as a parent band dispersion by the hybridization between the top and bottom edge states in three-dimensional TI [35,36,37]. A previous study suggests that a thickness of 5 QL is near the critical limit for inducing a hybridization gap in $Bi_2Se_3$ grown on 6H-SiC [35], supporting the plausibility of Rashba-type band splitting in the $S_5$ region of our $Bi_2Te_3/Cr_2Te_3$ system. A backscattering channel in a Rashba system generally does not necessarily imply time-reversal symmetry (TRS) breaking since two constant-energy contours (CECs) with different spin helicities coexist [38]. However, due to spectral weight transfer between the top and bottom surface, one of these CECs appears to be missing in our observations. This makes it difficult to determine whether the original electronic state is Dirac-based [**Fig. 6(d)**] or Rashba-based [**Fig. 6(e)**]. Regardless of whether the band structure follows helical Dirac or Rashba characteristics, our findings indicate that the observed gap opening and backscattering in the $S_5$ region are driven by TRS breaking due to magnetic perturbation, despite the absence of Cr diffusion in the 5$^{th}$ QL.

**Magnetic Perturbation Length Scale**

**Figures 6(f)** and **6(g)** schematically illustrate the long-range magnetic perturbation, considering both diffusion-driven and proximity effects (independent of Cr diffusion). In **Fig. 6(f)**, the gapped ($S_5$) and gapless ($S_6$) band regions are highlighted, corresponding to the shaded areas in **Figs. 6(d)–6(e)**. The length scale of the magnetic proximity effect (MPE) is typically measured relative to the parent magnetic material ($Cr_2Te_3$ in our case). From this perspective, the MPE extends approximately 5 nm. However, Cr gradually diffuses within the 1$^{st}$ to 4$^{th}$ QL region [see **Fig. 1(c)**]. If we instead define the MPE length scale as the region where it occurs without Cr diffusion, its characteristic length scale is around 1 nm—spanning the 4$^{th}$ to 5$^{th}$ QL. While this value roughly aligns with literature reports [12,13,14], the 4$^{th}$ QL should not be positioned the same as the parent magnetic material $Cr_2Te_3$. While it is a matter of terminology, in **Fig. 6(g)**, the region between the 1$^{st}$ and 4$^{th}$ QL is depicted as hosting both MPE and Cr diffusion, as it is not natural to consider an emerging MPE, which is disconnected from parent magnetic material $Cr_2Te_3$. Finally, we also note the intriguing viewpoint for the region with gradual Cr diffusion. Beyond the MPE itself, this region with broken inversion symmetry may also play a key role in realizing the exotic electronic and magnetic state of matters. For example, this region could host topologically nontrivial spin textures and Rashba bands with giant energy splitting [4,39,40]. Further investigation is valuable to explore the interplay between thickness-dependent gap evolution, gradual chemical diffusion, magnetic structure evolution, and spectral weight transfer between the top and bottom surfaces.

## V. SUMMARY

In summary, we demonstrate a long-range magnetic perturbation, reaching up to 5$^{th}$ QL in $Bi_2Te_3$ film grown on a thick ferromagnetic $Cr_2Te_3$ film. By harnessing chemical diffusion and the magnetic proximity effect, heterostructures integrating magnetic materials with material with high SOC (such as $Bi_2Te_3$ $Z_2$ topological insulators) enable long-range magnetic perturbations, offering a promising route for engineering topologically nontrivial electronic and magnetic states characterized in real and momentum spaces.

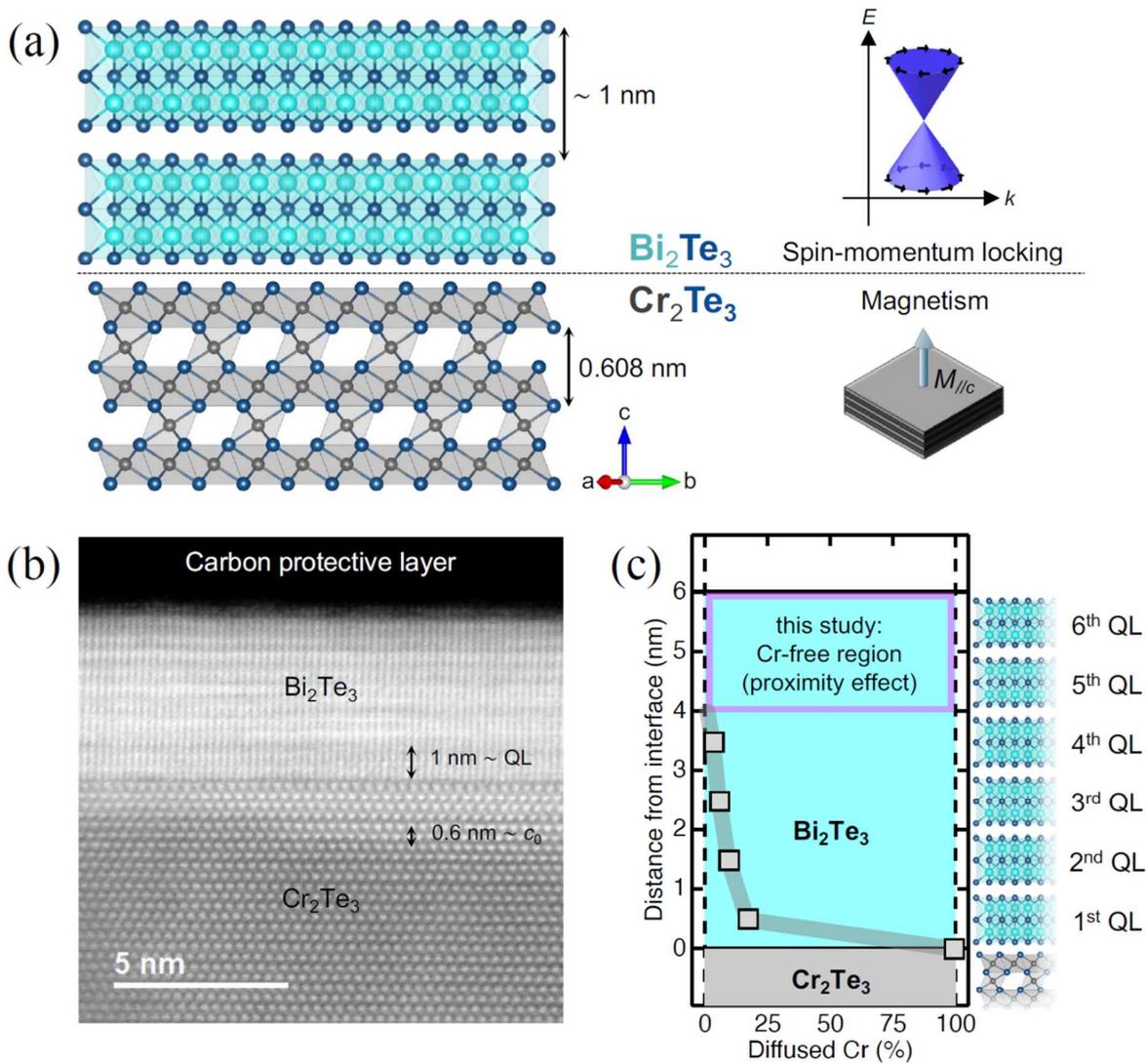

**FIG. 1. Structural and chemical characterization of the Bi$_2$Te$_3$/Cr$_2$Te$_3$ heterostructure.** (a) Schematic illustration of the magnetic TI heterostructure, highlighting the perpendicular magnetic anisotropy of the Cr$_2$Te$_3$ film ($M \parallel c$) and the spin-polarized Dirac cone of Bi$_2$Te$_3$. (b) Cross-sectional STEM image of the heterostructure, showing a 5 QL Bi$_2$Te$_3$ region. (c) Cr concentration profile in the Bi$_2$Te$_3$ film as a function of distance from the interface, revealing a Cr-free region above 4 nm (pink rectangle). Data points are extracted from STEM-EDS analysis (see Fig. S3 for details).

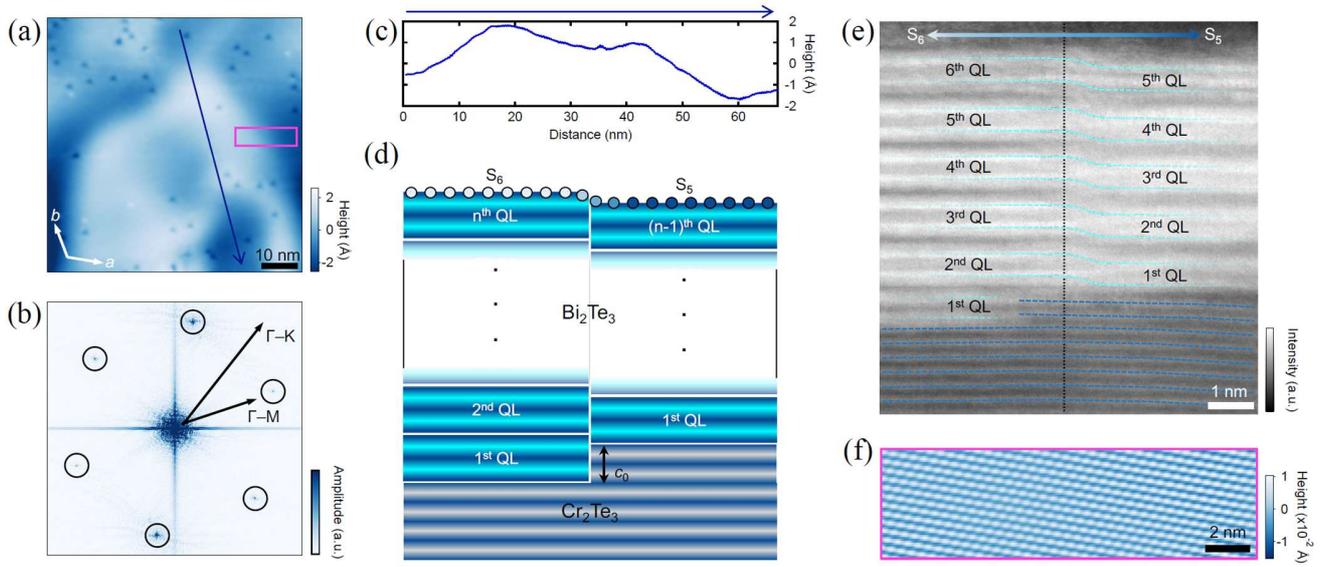

**FIG. 2. Surface morphology analysis via scanning tunneling microscope (STM) and scanning transmission electron microscope (STEM).** (a) STM topograph (70 × 70 nm) revealing ~4 Å corrugation on a continuous Te-terminated surface. The setup condition is 100 mV and 500 pA. (b) Fourier transform (FT) of (a), highlighting the primary crystallographic directions (arrows) and Bragg peaks (circles). (c) Line profile along the black line in (b), comparing the highest and lowest regions in the STM topography. (d) Schematic illustrating the connection between the observed surface corrugation in (b) and the distinct c-axis lattice parameters of $Cr_2Te_3$ and $Bi_2Te_3$ [see **Fig. 1(a)**, right]. (e) Cross-sectional STEM image showing the structural origin of surface corrugation between the S6 and S5 regions, attributed to the differing c parameters of $Cr_2Te_3$ and $Bi_2Te_3$. Dashed lines guide the eye for Te layers in $Cr_2Te_3$ (light blue) and Bi sites in $Bi_2Te_3$ (dark blue). (f) High-pass filtered STM images of $S_6$ and $S_5$ regions [see pink rectangle in (a)], reconstructed via inverse FT, illustrating the continuity of the topmost Te layer.

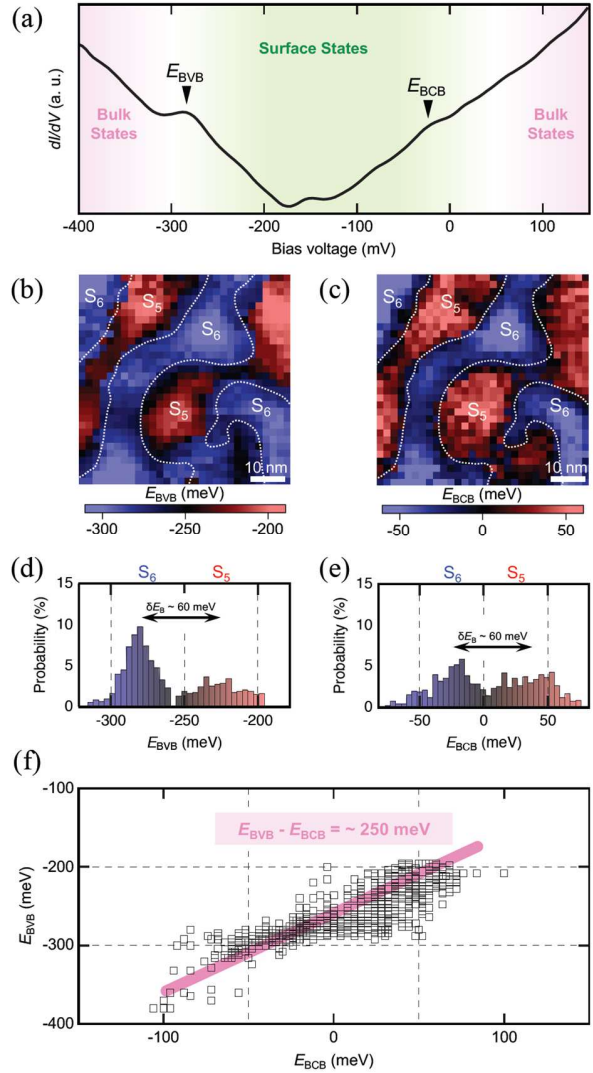

**FIG. 3. Spatial heterogeneity of the bulk valence and conduction bands in $Bi_2Te_3$.** (a) Spatially averaged $dI/dV$ spectra, indicating the bulk valence band maximum ($E_{BVB}$) and bulk conduction band minimum ($E_{BCB}$). The setup condition is -400 mV and 400 pA, with the lock-in amp setup of 10 mV bias modulation and a frequency of 961 Hz. (b), (c) Spatial maps of $E_{BVB}$ and $E_{BCB}$, respectively, corresponding to the terrace shown in **Fig. 2(a)**. The white dashed lines mark the smoothed boundary between the S5 and S6 regions, defined by the 50 % higher and 50 % lower energy points of $E_{BCB}$. (d), (e) Histograms of $E_{BVB}$ and $E_{BCB}$ for all data points in (b) and (c), revealing a bimodal distribution correlated with the $S_5$ and $S_6$ regions. (f) Scatter plot of $E_{BVB}(r)$ versus $E_{BCB}(r)$, showing a spatially constant bulk band gap of ~250 meV.

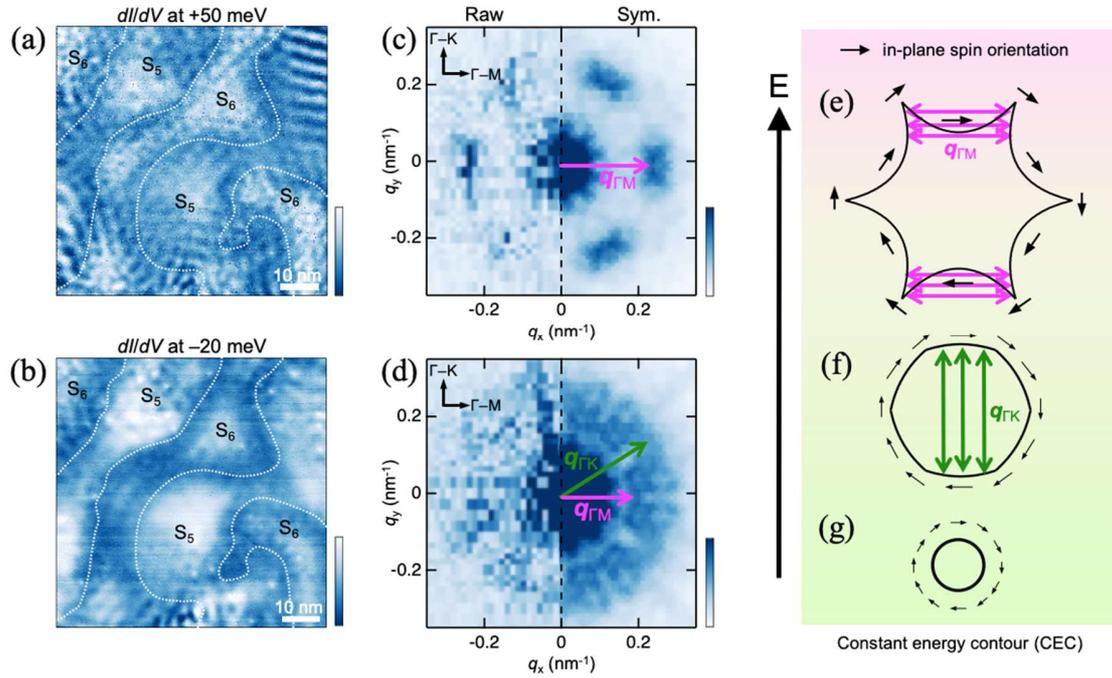

**FIG. 4. Spatially averaged quasiparticle interference of the surface state.** (a), (b) Constant energy contours (CECs) of $dI/dV$ maps at + 50 and - 20 meV, taken from the 70 × 70 nm terrace shown in **Fig. 2(a)**. The white dashed lines indicate the boundary between $S_5$ and $S_6$ regions [see **Fig. 3(c)**]. The setup condition is 100 mV and 500 pA, with the lock-in amp setup of 10 mV bias modulation and a frequency of 961 Hz. (c), (d) Raw (left) and symmetrized (right) Fourier transforms (FTs) of the CECs from (a) and (b), revealing six-fold quasiparticle interference (QPI) patterns along the Γ–M (pink) and Γ–K (green) directions, respectively. (e)-(g) CECs of $Bi_2Te_3$ at different energies (moving downward corresponds to approaching the Dirac point). The origin of $q_{ΓM}$ and $q_{ΓK}$ observed in (c) and (d) are indicated by pink and green arrows, respectively. Black arrows indicate spin orientation, adapted from [5,31].

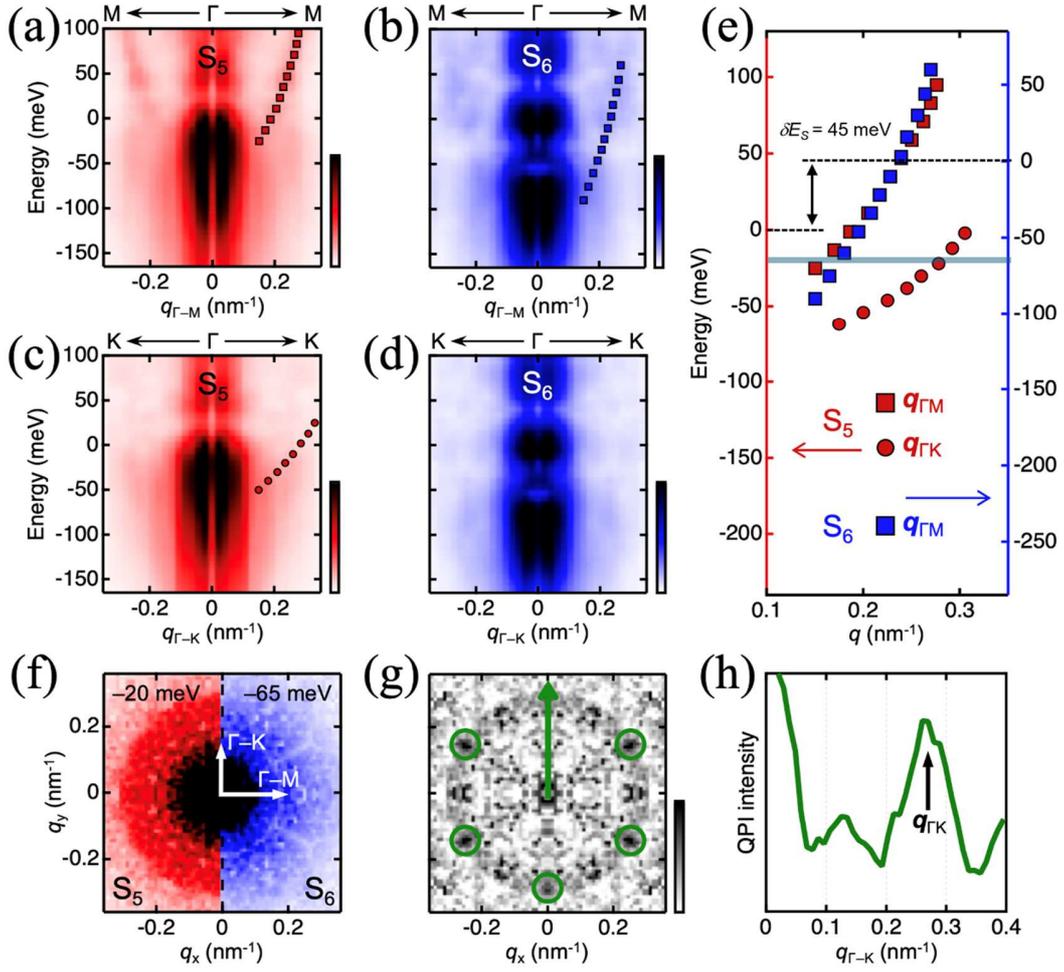

**FIG. 5. Comparison of energy-dependent quasiparticle interference (QPI) patterns between $S_5$ and $S_6$ regions.** (a)-(d) Energy evolution of QPI patterns along the Γ–M (a, b) and Γ–K (c, d) directions for S5 [red, (a, c)] and $S_6$ [blue, (b, d)] regions. Squares track the dispersion of $q_{ΓM}$, while circles track $q_{ΓK}$. (e) Extracted scattering vectors $q_{ΓM}(E)$ from (a) and (b). (f) $dI/dV$ FT CECs of $S_5$ (red, left) and $S_6$ (blue, right) regions at -20 and -65 meV, respectively. (g) The normalized QPI pattern obtained by dividing the $dI/dV$ FT CECs of the $S_5$ and $S_6$ regions, as shown in (f). Green circles highlight higher intensity regions due to enhanced QPI in the $S_5$ along Γ–K. (h) Line profile of the QPI intensity along the green arrow in (g), revealing $q_{ΓK}$.

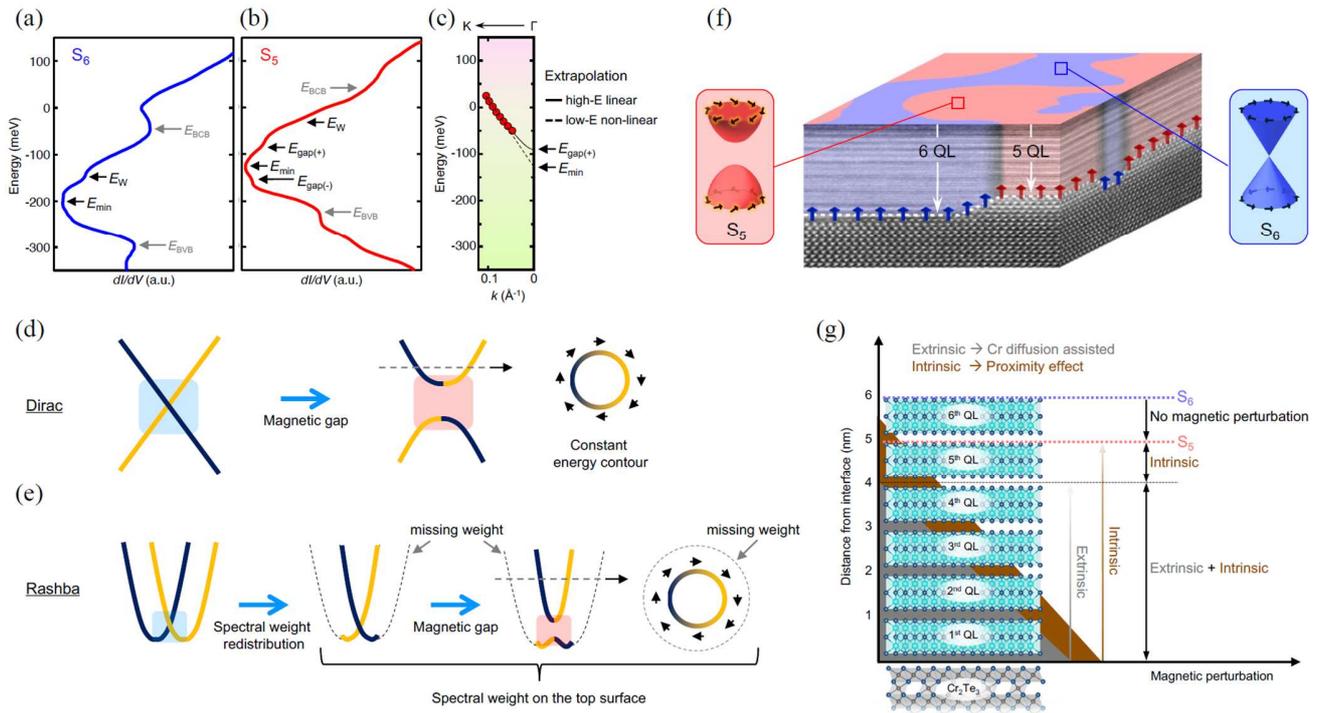

**FIG. 6. Collective interpretation of chemical, structural, and spectroscopic characterization in the Bi$_2$Te$_3$/Cr$_2$Te$_3$ heterostructure.** (a), (b) Spatially averaged $dI/dV$ spectra for the S$_5$ and S$_6$ regions, respectively. See the main text for details regarding the indicated features. The setup condition is -400 mV and 400 pA, with the lock-in amp setup of 10 mV bias modulation and a frequency of 961 Hz. (c) The momentum dispersion along ΓK direction, which is deduced from q$_{ΓK}$ from the S$_5$ region. The extrapolation to $k = 0$ limit allows to estimate E$_{gap(+)}$ (curved line) and charge neutral point $E_{min}$ (straight line). (d), (e) Schematic band structure of the top surface state. The Dirac (d) and Rashba (e) representations illustrate the parent electronic state in the S$_5$ region (see main text for details). (f) Schematic illustrating the correlation between cross-sectional STEM (lateral faces) data and bulk band electronic inhomogeneity (top surface), with gapped (S$_5$, red) and gapless (S$_6$, blue) regions. Black arrows denote the spin orientation of the TSS, which is distorted in the S$_5$ region due to time-reversal symmetry breaking. (g) Schematic for long-range magnetic perturbation, extending up to ~ 5 nm.